\documentclass[aip,reprint]{revtex4-1}
\usepackage{graphicx}
\usepackage{amsmath}
\usepackage{dcolumn} 
\usepackage{bm}

\usepackage[utf8]{inputenc}
\usepackage[T1]{fontenc}
\usepackage{mathptmx}
\usepackage{etoolbox}
\usepackage{makecell}

\draft 

\begin{document}

\title{Enhanced Current Density and Asymmetry of Metal-Insulator-Metal Diodes Based on the Self-Assembly of Pt Nanoparticles for Optical Rectennas} 

\author{Zhen Liu}
 \email[Corresponding author 1:] {	liu.zhen.p5@dc.tohoku.ac.jp}
\author{Shunsuke Abe}
\author{Makoto Shimizu}
 \email[Corresponding author 2:] {	makoto.shimizu.a3@tohoku.ac.jp}
\author{Hiroo Yugami}
\affiliation{Graduate School of Engineering, Tohoku University, Aoba 6-6-01 Aramaki, Aoba-ku, Sendai 980-8579, Japan}



\begin{abstract}
Optical rectennas consist of nano-antennas and nano-scale rectifying diodes, providing extensive prospects for thermal radiation energy harvesting applications. To achieve this, high current density and high asymmetry must be simultaneously obtained in rectifying diodes with ultra-high-speed responses to optical frequencies. In this study, we report a metal-insulator-metal (MIM) diode with a strongly enhanced electric field achieved via the self-assembly of uniform Pt nanoparticles (NPs) using atomic layer deposition. An enhancement of several orders of magnitude in the current density and asymmetry of this system in comparison to conventional MIM diodes was realized by shaping the tunneling barrier. The diode efficiency of the proposed MIM diodes experimentally confirmed that significantly exceeds the MIM diode without NPs by 231 times. Furthermore, the proposed strategy can be integrated with various advanced tunnel diodes to achieve high-performance optical rectennas.
\end{abstract}

\pacs{}

\maketitle 


Optical rectennas consist of an antenna-coupled rectifying element that directly converts electromagnetic waves into electrical energy. These systems have attracted widespread attention due to their super-wide operating frequencies and potentially high energy-conversion efficiency\cite{JAYASWAL20181,Matsuura2022}. A variety of diode devices have been investigated for use in infrared radiation rectification\cite{Matarrese1970,Twu1974}, infrared (IR) detection\cite{Gustafson1974,Hobbs:05,Bean2011}, and IR imaging\cite{Grover2010,https://doi.org/10.1002/mop.27363}. Metal-insulator-metal (MIM) tunnel diodes are strong candidates for use as rectifiers in optical rectennas due to their high-speed response toward ultra-fast pole changes within the optical frequency range from the visible to IR frequencies. The tunneling-based optical rectification mechanism allows electron tunneling in the MIM junction on the order of femtoseconds. Therefore, the nanoscale MIM diodes attract increasing attention to use in optical rectennas which contribute to energy harvesting techniques including solar energy conversion, waste-heat utilization, and environmental thermal radiation\cite{Grover2011,Sharma2015,Dragoman2016,Herner2017,Anderson2019,doi:10.1126/science.aba2089,Elsharabasy2020}.

Low resistivity is essential to obtain a low RC time constant and hence a high current density at high cutoff frequencies. Moreover, a high rectification performance is obtained in terms of the large current density asymmetry at forward and reverse biases. However, the tunnel current is exponentially and inversely dependent upon the barrier height and insulator thickness\cite{Periasamy2013,8272341}. The presence of asymmetry is a nonlinear response of diodes within the forward and reverse biases due to the work function difference. Several studies have been reported to overcome the trade-off relationship between the current density and asymmetry\cite{Tekin2021,Maraghechi2012,Alimardani2013,Cui2016}. Belkadi \emph{et al.} first observed resonant tunneling behavior in MIM diodes that have two different insulators,wherein a resistance reduction was observed and the responsivity increased\cite{Belkadi2021}. Matsuura \emph{et al.} demonstrated the improvement of MIM diodes by considerable orders of magnitude that consisted of combinations of oxygen stoichiometry and non-stoichiometry insulators\cite{Matsuura2019}. Moreover, geometrically asymmetric MIM diodes provide an appealing strategy because of the enhancement of the electric field\cite{Ward2010,Choi2011,Shin2016,Piltan2017,PhysRevB.96.115435,Shin2017}. However, the electric field concentration is strongly dependent on the geometrical structure of the electrodes and the precision of fabrication, which makes it challenging to scale up MIM diodes for low-cost energy harvesting applications.

In this study, we propose a novel tunnel diode in which metal nanoparticles (NPs) are formed at interface of the tunneling layer and electrode to implement an electric field concentration effect. Simulations of the electric field distribution in this structure are used to verify the strong electric field concentration at the NPs, and an analysis of the diode properties considering the variation of the tunneling barrier shape due to the electric field concentration is conducted to investigate the increase in current density and expression of asymmetry. This effect is examined by evaluating a tunneling diode in which NPs are formed by controlling the initial process of atomic layer deposition (ALD). The self-assembly of NPs more easily provides an electric field concentration effect in the tunneling diode comparing with previously developed methods. Furthermore, the proposed approach is not limited by materials and electrode geometries, which indicates this system has the potential to be applied towards the design of high-performance MIM diodes.

\begin{figure}[htbp]
\includegraphics[scale=0.6]{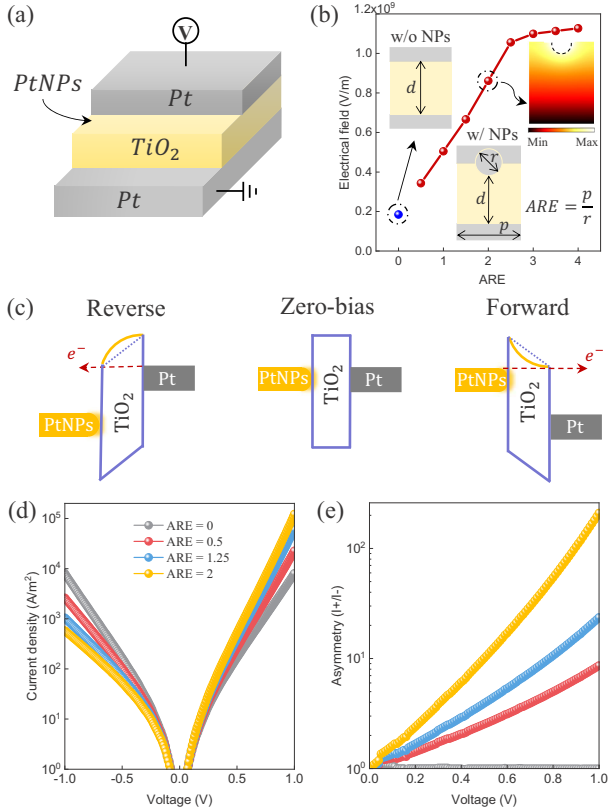}
\caption{(a) Schematic illustration of the PtNPs/ TiO2/Pt MIM diode with uniform NPs obtained using ALD. (b) Maximum electric field of the MIM diodes dependents on the geometrical parameter ARE. The inset shows the cross-section of a unit cell; all key geometrical parameters are indicated. The heatmap shows the potential distribution at ARE=2. (c) Schematic illustration of the band diagrams for the PtNPs/TiO\textsubscript{2}/Pt MIM diode under the electric field concentration effect. Calculated (d) current density and (e) asymmetry of the PtNPs/TiO\textsubscript{2}/Pt MIM diode.}
\label{fig_1}
\end{figure}

A schematic illustration of a NPs-containing MIM diode with a vertically stacked structure is shown in Fig. \ref{fig_1}(a). The Pt NPs were distributed on the interface of TiO\textsubscript{2} insulator and sandwiched Pt/Pt metal electrodes to form PtNPs/TiO\textsubscript{2}/Pt MIM diode. Pt NPs contribute to the electric field concentration by forming geometrical electrodes within entire contact interface of the top electrode. The electric field distribution of the unit cell from MIM diode containing NPs was analyzed using the Ansys Maxwell electromechanical simulation tool. The asymmetric ratio of electrodes (ARE) is defined as the ratio of the pitch size $p$ to the diameter of NPs $r$ to describe the structural features of the asymmetric electrodes. As shown in Fig. \ref{fig_1}(b), the electric field enhancement and ARE demonstrate an almost proportional relationship until ARE=2.5. The potential distribution in heatmap verify the strong electric field concentrated at the NPs. We note that the magnitude enhancement of electric field is attributed to the large ARE, which indicates that the size-controlled NPs are required to obtain a strong electric field concentration.

The potential barrier shape was determined using the effective work function $\varphi$=5.3 eV for both the PtNPs and Pt electrodes, which are based on the crystallographic orientation of electrodes according to results obtained via XRD and literature\cite{Gu2006}. The electron affinity $\chi$=3.9 eV and relative dielectric constant $\kappa$=10 for TiO\textsubscript{2} were obtained from literature\cite{Cui2016}. A schematic illustration of the band diagrams of the PtNPs/TiO\textsubscript{2}/Pt MIM diode at different biases is shown in Fig. \ref{fig_1}(c). The same barrier height at the zero-bias indicates that the effect of the work function difference has been eliminated, and a uniform electron density distribution is present in the tunnel layer. Upon the application of a voltage to the electrodes, the electron density becomes condensed at the PtNPs electrode side. The effective width of the tunnel barrier becomes shorter in the forward bias and longer in the reverse bias directions, and the tunneling probability of electrons is subsequently improved or suppressed in the corresponding bias direction, respectively. The shaping of the potential barrier induced by the electric field concentration effect was utilized to analyze the IV characteristics of this system via the WKB approximation method\cite{GROVER201294}. To clarify the effect of electric field enhancement, the IV characteristics of various AREs were investigated with 5.5-nm-thick of TiO\textsubscript{2} insulator. The effective mass of electrons $m^*$=0.16$\times m$ was determined based on the insulator thickness according to literature\cite{Cui2016}. The calculated current density and asymmetry of the PtNPs/TiO\textsubscript{2}/Pt MIM diodes are shown in Figs. \ref{fig_1}(d) and (e), respectively. The improvement of the rectification performance is consistent with the analysis of the tunneling barrier shaping process considering the electric field concentration effect. Therefore, the PtNPs/TiO\textsubscript{2}/Pt MIM diode should exhibit a large asymmetry.

The PtNPs/TiO\textsubscript{2}/Pt MIM diode was fabricated on a quartz substrate. A 70-nm-thick Pt layer was deposited by radio frequency magnetron sputtering (Shibaura, CEF-4ES), preceded by an 8-nm-thick titanium adhesion layer. A photoresist was applied and patterned via lithography (SUSS, Microtec MA6) and the bottom of the Pt electrodes on the MIM diode was formed by ion beam milling (Hakuto, IBE-KDC 75). A 5.5-nm-thick TiO\textsubscript{2} layer was deposited by ALD at 190 $^{\circ}$C. The Pt NPs were grown on the surface of TiO\textsubscript{2} at 275 $^{\circ}$C over 20 cycles. Trimethyl-methylcyclopentadienyl platinum and oxygen were used as the precursor and reaction gas, respectively. A 0.5-nm-thick TiO\textsubscript{2} layer was deposited onto Pt NPs by ALD at 190 $^{\circ}$C to ensure the formation of the array of separate NP islands. Finally, a 70-nm-thick Pt film was deposited as the top electrode under the same conditions used for the formation of the bottom electrode. A schematic illustration of the fabrication process is shown in Fig. S1.

\begin{figure}[htbp]
\includegraphics[scale=0.6]{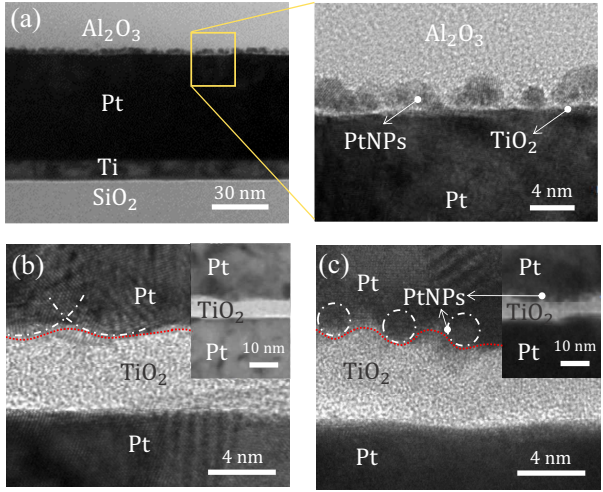}
\caption{Cross sectional TEM images of the (a) arrays of Pt NP islands grown on the substrate, (b) Pt/TiO\textsubscript{2}/Pt MIM diode without NPs, and (c) PtNPs/TiO\textsubscript{2}/Pt MIM diode with NPs. The TiO\textsubscript{2} insulator was deposited over 100 cycles using ALD.}
\label{fig_2}
\end{figure}

The growth behavior of nanometals deposited via ALD has well researched\cite{Juppo2000,Elam2003}. Uniform NPs are synthesized via island-growth at the initial stage, as described by the Volmer-Weber growth mode\cite{Shrestha_2010}. The excellent self-assembly nucleation of Pt in ALD has been reported in several studies\cite{Pyeon2015,Shimizu2016b,https://doi.org/10.1002/admi.201901210}. We successfully fabricated arrays of island Pt NPs using ALD and confirmed this deposition using transmission electrode microscopy (TEM)[Fig. \ref{fig_2}(a)]. The Pt was grown by ALD over 20 cycles on the substrate and Al\textsubscript{2}O\textsubscript{3} was coated as the protection layer. The self-assembled array of Pt NPs was observed in the magnified images. As shown in Figs. \ref{fig_2}(b) and \ref{fig_2}(c), the estimated boundaries between the top electrode and insulator are depicted by the red dotted lines. The wavy outline of the upper interface was clearly observed in the MIM diode containing NPs, which demonstrates that geometrical electrodes were formed by embedding the uniform Pt NPs arrays in the tunneling layer. However, small roughness was observed in the MIM diodes without NPs. We infer that the non-perfect plane interface was generated due to the grain growth of Pt during the deposition of the top electrode. The upper interfacial boundaries were fitted using a circular arc to analyze the geometric parameters of the electrodes, which are shown by the white dashed lines. The ARE can be estimated using the ratio of the pitch and diameter of the fitted circular arc. Using these techniques, the geometrical structural parameters were extracted from the TEM images to model the MIM diodes with the electric field concentration effect.

\begin{figure}[htbp]
\includegraphics[scale=0.3]{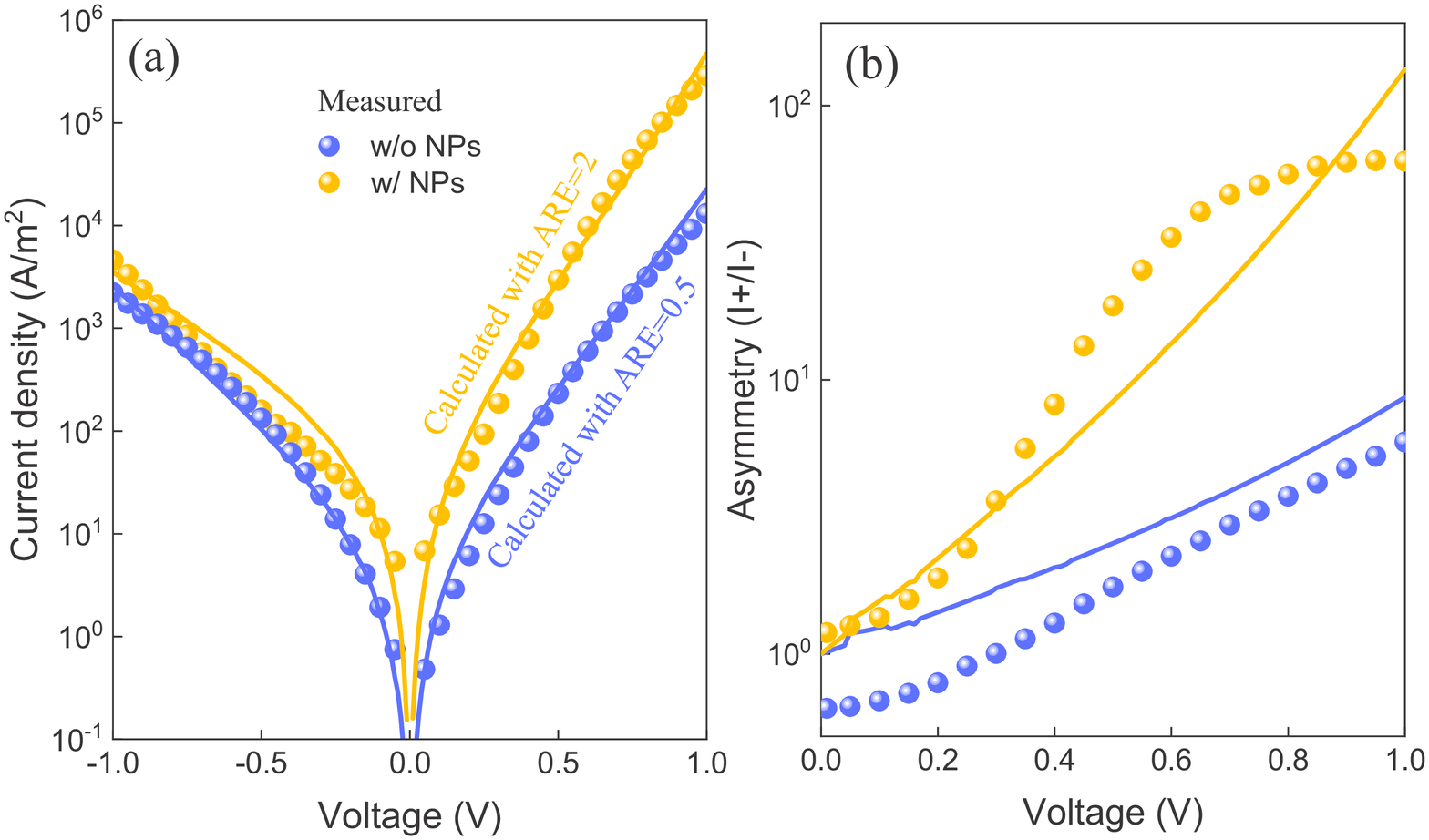}
\caption{(a) Current density and (b) asymmetry versus the operating voltage for the MIM diodes with and without NPs. The dots show the measured results obtained using the fabricated MIM diode, and the solid lines show the calculated results obtained using the MIM tunnel barrier models with the electric field enhancement effect.}
\label{fig_3}
\end{figure}

The plots of current density and asymmetry versus the voltage for the diodes with and without PtNPs are shown in Fig. \ref{fig_3}. The IV characteristics were measured at room temperature using the darkroom with a source measurement unit (Keithley, 2461-900-02A). The IV curves were swept from -1 to +1 V to demonstrate the potential of this system for energy harvesting applications. The measured and calculated results are plotted by the dotted and solid lines, respectively. An asymmetry of 5.9 at 1 V was obtained for the Pt/TiO\textsubscript{2}/Pt MIM diode. This asymmetry may result from the weak electric field concentration effect, which was induced from the roughness of the upper interface. An ARE=0.5 and the presence of a 5.5-nm-thick insulator were estimated from Fig. \ref{fig_2}(b), and these parameters used to model the variation of the barrier shape during electron tunneling. The calculated IV-curve showed a good agreement with the experimental results, which indicates the existence of a weak electric field enhancement in this MIM diode without NPs. A maximum current density of $2.9\times 10^5$ A/m$^2$ and asymmetry of 63 at 1 V were observed for the PtNPs/TiO\textsubscript{2}/Pt MIM diode. Improvements of 21- and 11-times were observed for the current density and asymmetry of this diode, respectively, compared to the MIM diode without NPs. The significant improvement is attributed to the controlling of the electron tunneling probability, which results from the electric field concentration effect achieved by embedding NPs in the tunneling layer as geometric electrodes. The MIM diode with NPs was modelled using geometrical parameters of ARE=2 and a 5.2-nm-thick insulator, which were extracted from Fig. \ref{fig_2}(c). Generally, the calculated results are shown to be consistent with the measured results in terms of the current density and asymmetry improvements. These results serve to verify the strong electric field concentration effect in the PtNPs/TiO\textsubscript{2}/Pt MIM diode. The slight deviation of the asymmetry expression from the measured results may have occurred because the self-assembly of nanometals via ALD can not provide perfectly periodic arrays of NPs. Therefore, the electric field distribution that strongly depends on roughness is complicated in the presence of electron tunneling\cite{Gaillard2006}. In this study, the electric field enhancement was utilized to demonstrate the performance improvement of the MIM diode containing NPs. The numerical analysis of MIM diodes that are formed using self-assembled nanometals via ALD will conduct in a further study.

\begin{figure}[htbp]
\includegraphics[scale=0.55]{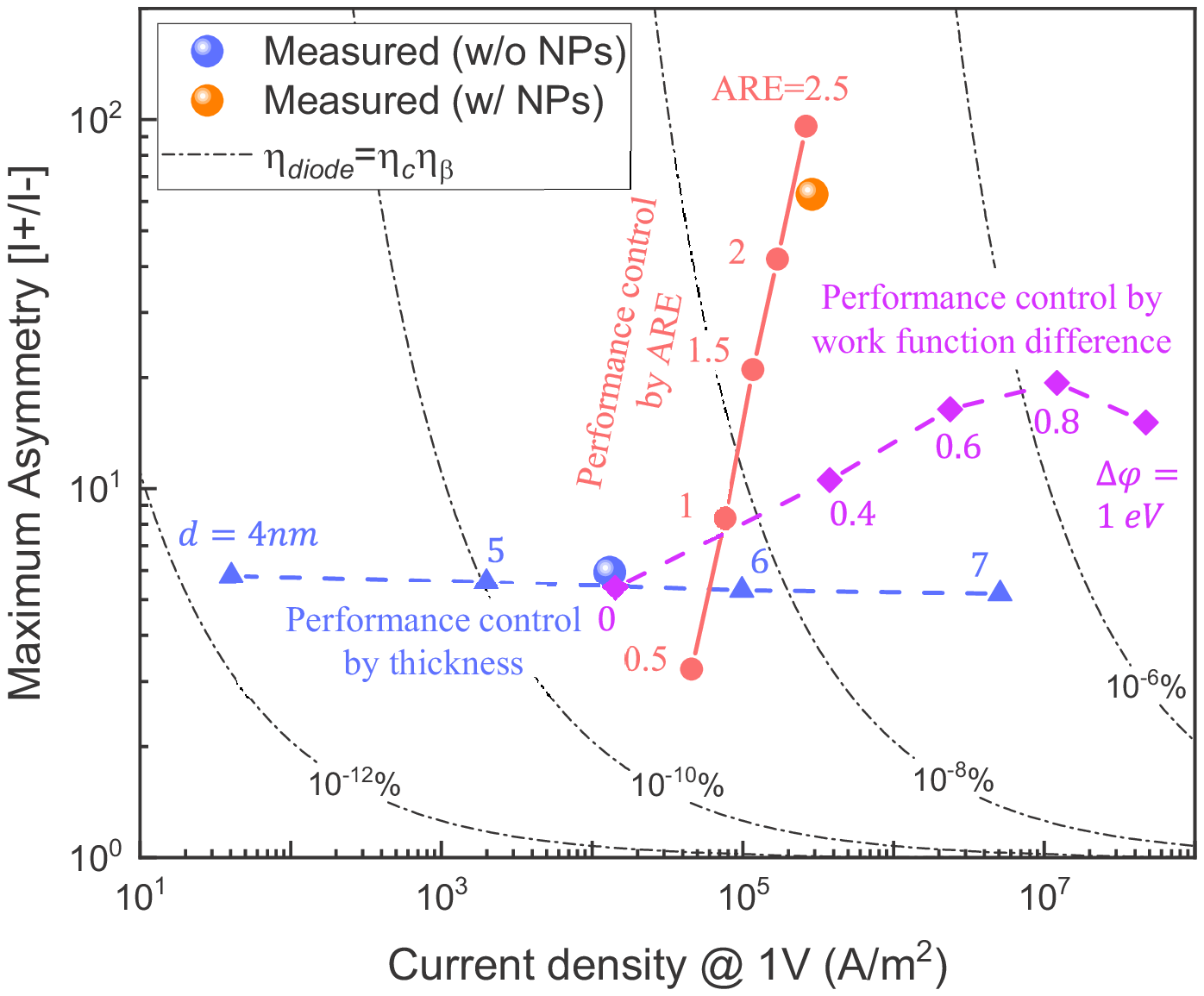}
\caption{Calculated performances of the MIM diodes at 1 V. The dashed lines indicate the performances control using the conventional approaches with the simulation model of an MIM diode without NPs, wherein the thickness is varied from 4 to 7 nm with a work function difference of $\Delta\varphi$=0 eV (blue) and work function difference from 0 to 1 eV with a thickness of $d=5.5$ nm (purple). The solid line indicates the performance of an MIM diode with NPs and a 5.5-nm-thick insulator with an ARE of 0.5 to 2.5.}
\label{fig_4}
\end{figure}

The diode performance in the optical rectenna systems can be evaluated by considering the current density and maximum asymmetry. This evaluation method was demonstrated in our previous study, which depicted the total diode efficiency $\eta_{diode}$ as being determined by the impedance matching efficiency $\eta_{c}$ and quantum efficiency $\eta_{\beta}$\cite{Matsuura2019}. The performances of the proposed MIM diodes are shown in Fig. \ref{fig_4}, where the measured results of the MIM diode with NPs was improved by 231 times compared to the MIM diode without NPs. The calculated performance of the MIM diode without NPs reveals a clear trade-off relationship experienced when using the conventional approach to control the insulator thickness. Besides, the approach of increasing the work function difference of diodes only provides a limited increase in asymmetry. However, the MIM diode with NPs achieved considerable asymmetry as the electric field concentration effect was enhanced by the ARE increased. In addition, the effect of the work function difference was not considered in the performance calculations to clarify the effect of the electric field concentration. Therefore, the diode efficiency of this system is expected to exhibit a higher improvement of significant orders of magnitude which relies on the low barrier height and different work functions of the electrode materials in this system.

In this study, we proposed a novel MIM diode with metal NPs that was grown via the self-assembly of Pt NPs at the interface of insulator and electrodes over the entire diode plane. The embedded Pt NPs serve as geometrical electrodes to achieve an electric field concentration effect and significantly enhance the current density and asymmetry simultaneously in this system. The performance improvement of the MIM diode with NPs was analyzed by considering the effect of the electric field enhancement. PtNPs/TiO\textsubscript{2}/Pt MIM diode was fabricated by coating a substrate with size-controlled island-grown Pt NPs via ALD, and the geometrical features of the structure were confirmed using TEM images. The performance of the MIM diode with NPs was experimentally confirmed and demonstrated an improvement of 231 times compared to the MIM diode fabricated without NPs. The improvement of the tunneling probability in the MIM diode with NPs due to the electric field concentration effect resulting from the asymmetric electrodes significantly exceeds the theoretical performance of conventional MIM diodes. Furthermore, this electric field concentration effect achieved by self-assembling NPs can universally apply in various MIM diodes. Improving the efficiency of diodes for energy harvesting purposes is expected by integrating this approach with well-designed high-frequency-response tunnel diodes.\\

See the supplementary material for the fabrication detail of PtNPs/TiO\textsubscript{2}/Pt MIM diode.\\

We would like to gratefully acknowledge that this paper is based on results of obtained from a project commissioned by the New Energy and Industrial Technology Development Organization (NEDO). Part of this work was supported by the Japan Society for the Promotion of Science (JSPS) (Grant Number: 22J11167).

\bibliographystyle{apsrev4-1}
\bibliography{reference}

\begin{thebibliography}{39}%
\makeatletter
\providecommand \@ifxundefined [1]{%
 \@ifx{#1\undefined}
}%
\providecommand \@ifnum [1]{%
 \ifnum #1\expandafter \@firstoftwo
 \else \expandafter \@secondoftwo
 \fi
}%
\providecommand \@ifx [1]{%
 \ifx #1\expandafter \@firstoftwo
 \else \expandafter \@secondoftwo
 \fi
}%
\providecommand \natexlab [1]{#1}%
\providecommand \enquote  [1]{``#1''}%
\providecommand \bibnamefont  [1]{#1}%
\providecommand \bibfnamefont [1]{#1}%
\providecommand \citenamefont [1]{#1}%
\providecommand \href@noop [0]{\@secondoftwo}%
\providecommand \href [0]{\begingroup \@sanitize@url \@href}%
\providecommand \@href[1]{\@@startlink{#1}\@@href}%
\providecommand \@@href[1]{\endgroup#1\@@endlink}%
\providecommand \@sanitize@url [0]{\catcode `\\12\catcode `\$12\catcode
  `\&12\catcode `\#12\catcode `\^12\catcode `\_12\catcode `\%12\relax}%
\providecommand \@@startlink[1]{}%
\providecommand \@@endlink[0]{}%
\providecommand \url  [0]{\begingroup\@sanitize@url \@url }%
\providecommand \@url [1]{\endgroup\@href {#1}{\urlprefix }}%
\providecommand \urlprefix  [0]{URL }%
\providecommand \Eprint [0]{\href }%
\providecommand \doibase [0]{http://dx.doi.org/}%
\providecommand \selectlanguage [0]{\@gobble}%
\providecommand \bibinfo  [0]{\@secondoftwo}%
\providecommand \bibfield  [0]{\@secondoftwo}%
\providecommand \translation [1]{[#1]}%
\providecommand \BibitemOpen [0]{}%
\providecommand \bibitemStop [0]{}%
\providecommand \bibitemNoStop [0]{.\EOS\space}%
\providecommand \EOS [0]{\spacefactor3000\relax}%
\providecommand \BibitemShut  [1]{\csname bibitem#1\endcsname}%
\let\auto@bib@innerbib\@empty
\bibitem [{\citenamefont {Jayaswal}\ \emph {et~al.}(2018)\citenamefont
  {Jayaswal}, \citenamefont {Belkadi}, \citenamefont {Meredov}, \citenamefont
  {Pelz}, \citenamefont {Moddel},\ and\ \citenamefont
  {Shamim}}]{JAYASWAL20181}%
  \BibitemOpen
  \bibfield  {author} {\bibinfo {author} {\bibfnamefont {G.}~\bibnamefont
  {Jayaswal}}, \bibinfo {author} {\bibfnamefont {A.}~\bibnamefont {Belkadi}},
  \bibinfo {author} {\bibfnamefont {A.}~\bibnamefont {Meredov}}, \bibinfo
  {author} {\bibfnamefont {B.}~\bibnamefont {Pelz}}, \bibinfo {author}
  {\bibfnamefont {G.}~\bibnamefont {Moddel}}, \ and\ \bibinfo {author}
  {\bibfnamefont {A.}~\bibnamefont {Shamim}},\ }\href {\doibase
  https://doi.org/10.1016/j.mtener.2017.11.002} {\bibfield  {journal} {\bibinfo
   {journal} {Materials Today Energy}\ }\textbf {\bibinfo {volume} {7}},\
  \bibinfo {pages} {1} (\bibinfo {year} {2018})}\BibitemShut {NoStop}%
\bibitem [{\citenamefont {Matsuura}\ \emph {et~al.}(2022)\citenamefont
  {Matsuura}, \citenamefont {Shimizu}, \citenamefont {Liu},\ and\ \citenamefont
  {Yugami}}]{Matsuura2022}%
  \BibitemOpen
  \bibfield  {author} {\bibinfo {author} {\bibfnamefont {D.}~\bibnamefont
  {Matsuura}}, \bibinfo {author} {\bibfnamefont {M.}~\bibnamefont {Shimizu}},
  \bibinfo {author} {\bibfnamefont {Z.}~\bibnamefont {Liu}}, \ and\ \bibinfo
  {author} {\bibfnamefont {H.}~\bibnamefont {Yugami}},\ }\href@noop {}
  {\bibfield  {journal} {\bibinfo  {journal} {Applied Physics Express}\
  }\textbf {\bibinfo {volume} {15}} (\bibinfo {year} {2022})}\BibitemShut
  {NoStop}%
\bibitem [{\citenamefont {Matarrese}\ and\ \citenamefont
  {Evenson}(1970)}]{Matarrese1970}%
  \BibitemOpen
  \bibfield  {author} {\bibinfo {author} {\bibfnamefont {L.~M.}\ \bibnamefont
  {Matarrese}}\ and\ \bibinfo {author} {\bibfnamefont {K.~M.}\ \bibnamefont
  {Evenson}},\ }\href {\doibase 10.1063/1.1653250} {\bibfield  {journal}
  {\bibinfo  {journal} {Applied Physics Letters}\ }\textbf {\bibinfo {volume}
  {17}},\ \bibinfo {pages} {8} (\bibinfo {year} {1970})}\BibitemShut {NoStop}%
\bibitem [{\citenamefont {Twu}\ and\ \citenamefont {Schwarz}(1974)}]{Twu1974}%
  \BibitemOpen
  \bibfield  {author} {\bibinfo {author} {\bibfnamefont {B.}~\bibnamefont
  {Twu}}\ and\ \bibinfo {author} {\bibfnamefont {S.~E.}\ \bibnamefont
  {Schwarz}},\ }\href {\doibase 10.1063/1.1655325} {\bibfield  {journal}
  {\bibinfo  {journal} {Applied Physics Letters}\ }\textbf {\bibinfo {volume}
  {25}},\ \bibinfo {pages} {595} (\bibinfo {year} {1974})}\BibitemShut
  {NoStop}%
\bibitem [{\citenamefont {Gustafson}\ \emph {et~al.}(1974)\citenamefont
  {Gustafson}, \citenamefont {Schmidt},\ and\ \citenamefont
  {Perucca}}]{Gustafson1974}%
  \BibitemOpen
  \bibfield  {author} {\bibinfo {author} {\bibfnamefont {T.~K.}\ \bibnamefont
  {Gustafson}}, \bibinfo {author} {\bibfnamefont {R.~V.}\ \bibnamefont
  {Schmidt}}, \ and\ \bibinfo {author} {\bibfnamefont {J.~R.}\ \bibnamefont
  {Perucca}},\ }\href {\doibase 10.1063/1.1655078} {\bibfield  {journal}
  {\bibinfo  {journal} {Applied Physics Letters}\ }\textbf {\bibinfo {volume}
  {24}},\ \bibinfo {pages} {620} (\bibinfo {year} {1974})}\BibitemShut
  {NoStop}%
\bibitem [{\citenamefont {Hobbs}\ \emph {et~al.}(2005)\citenamefont {Hobbs},
  \citenamefont {Laibowitz},\ and\ \citenamefont {Libsch}}]{Hobbs:05}%
  \BibitemOpen
  \bibfield  {author} {\bibinfo {author} {\bibfnamefont {P.~C.~D.}\
  \bibnamefont {Hobbs}}, \bibinfo {author} {\bibfnamefont {R.~B.}\ \bibnamefont
  {Laibowitz}}, \ and\ \bibinfo {author} {\bibfnamefont {F.~R.}\ \bibnamefont
  {Libsch}},\ }\href {\doibase 10.1364/AO.44.006813} {\bibfield  {journal}
  {\bibinfo  {journal} {Applied Optics}\ }\textbf {\bibinfo {volume} {44}},\
  \bibinfo {pages} {6813} (\bibinfo {year} {2005})}\BibitemShut {NoStop}%
\bibitem [{\citenamefont {Bean}\ \emph {et~al.}(2011)\citenamefont {Bean},
  \citenamefont {Weeks},\ and\ \citenamefont {Boreman}}]{Bean2011}%
  \BibitemOpen
  \bibfield  {author} {\bibinfo {author} {\bibfnamefont {J.~A.}\ \bibnamefont
  {Bean}}, \bibinfo {author} {\bibfnamefont {A.}~\bibnamefont {Weeks}}, \ and\
  \bibinfo {author} {\bibfnamefont {G.~D.}\ \bibnamefont {Boreman}},\ }\href
  {\doibase 10.1109/JQE.2010.2081971} {\bibfield  {journal} {\bibinfo
  {journal} {IEEE Journal of Quantum Electronics}\ }\textbf {\bibinfo {volume}
  {47}},\ \bibinfo {pages} {126} (\bibinfo {year} {2011})}\BibitemShut
  {NoStop}%
\bibitem [{\citenamefont {Grover}\ \emph {et~al.}(2010)\citenamefont {Grover},
  \citenamefont {Dmitriyeva}, \citenamefont {Estes},\ and\ \citenamefont
  {Moddel}}]{Grover2010}%
  \BibitemOpen
  \bibfield  {author} {\bibinfo {author} {\bibfnamefont {S.}~\bibnamefont
  {Grover}}, \bibinfo {author} {\bibfnamefont {O.}~\bibnamefont {Dmitriyeva}},
  \bibinfo {author} {\bibfnamefont {M.~J.}\ \bibnamefont {Estes}}, \ and\
  \bibinfo {author} {\bibfnamefont {G.}~\bibnamefont {Moddel}},\ }\href
  {\doibase 10.1109/TNANO.2010.2051334} {\bibfield  {journal} {\bibinfo
  {journal} {IEEE Transactions on Nanotechnology}\ }\textbf {\bibinfo {volume}
  {9}},\ \bibinfo {pages} {716} (\bibinfo {year} {2010})}\BibitemShut {NoStop}%
\bibitem [{\citenamefont {Kinzel}\ \emph {et~al.}(2013)\citenamefont {Kinzel},
  \citenamefont {Brown}, \citenamefont {Ginn}, \citenamefont {Lail},
  \citenamefont {Slovick},\ and\ \citenamefont
  {Boreman}}]{https://doi.org/10.1002/mop.27363}%
  \BibitemOpen
  \bibfield  {author} {\bibinfo {author} {\bibfnamefont {E.~C.}\ \bibnamefont
  {Kinzel}}, \bibinfo {author} {\bibfnamefont {R.~L.}\ \bibnamefont {Brown}},
  \bibinfo {author} {\bibfnamefont {J.~C.}\ \bibnamefont {Ginn}}, \bibinfo
  {author} {\bibfnamefont {B.~A.}\ \bibnamefont {Lail}}, \bibinfo {author}
  {\bibfnamefont {B.~A.}\ \bibnamefont {Slovick}}, \ and\ \bibinfo {author}
  {\bibfnamefont {G.~D.}\ \bibnamefont {Boreman}},\ }\href {\doibase
  https://doi.org/10.1002/mop.27363} {\bibfield  {journal} {\bibinfo  {journal}
  {Microwave and Optical Technology Letters}\ }\textbf {\bibinfo {volume}
  {55}},\ \bibinfo {pages} {489} (\bibinfo {year} {2013})}\BibitemShut
  {NoStop}%
\bibitem [{\citenamefont {Grover}\ and\ \citenamefont
  {Moddel}(2011)}]{Grover2011}%
  \BibitemOpen
  \bibfield  {author} {\bibinfo {author} {\bibfnamefont {S.}~\bibnamefont
  {Grover}}\ and\ \bibinfo {author} {\bibfnamefont {G.}~\bibnamefont
  {Moddel}},\ }\href {\doibase 10.1109/JPHOTOV.2011.2160489} {\bibfield
  {journal} {\bibinfo  {journal} {IEEE Journal of Photovoltaics}\ }\textbf
  {\bibinfo {volume} {1}},\ \bibinfo {pages} {78} (\bibinfo {year}
  {2011})}\BibitemShut {NoStop}%
\bibitem [{\citenamefont {Sharma}\ \emph {et~al.}(2015)\citenamefont {Sharma},
  \citenamefont {Singh}, \citenamefont {Bougher},\ and\ \citenamefont
  {Cola}}]{Sharma2015}%
  \BibitemOpen
  \bibfield  {author} {\bibinfo {author} {\bibfnamefont {A.}~\bibnamefont
  {Sharma}}, \bibinfo {author} {\bibfnamefont {V.}~\bibnamefont {Singh}},
  \bibinfo {author} {\bibfnamefont {T.~L.}\ \bibnamefont {Bougher}}, \ and\
  \bibinfo {author} {\bibfnamefont {B.~A.}\ \bibnamefont {Cola}},\ }\href
  {\doibase 10.1038/nnano.2015.220} {\bibfield  {journal} {\bibinfo  {journal}
  {Nature Nanotechnology}\ }\textbf {\bibinfo {volume} {10}},\ \bibinfo {pages}
  {1027} (\bibinfo {year} {2015})}\BibitemShut {NoStop}%
\bibitem [{\citenamefont {Dragoman}\ and\ \citenamefont
  {Aldrigo}(2016)}]{Dragoman2016}%
  \BibitemOpen
  \bibfield  {author} {\bibinfo {author} {\bibfnamefont {M.}~\bibnamefont
  {Dragoman}}\ and\ \bibinfo {author} {\bibfnamefont {M.}~\bibnamefont
  {Aldrigo}},\ }\href {\doibase 10.1063/1.4962642} {\bibfield  {journal}
  {\bibinfo  {journal} {Applied Physics Letters}\ }\textbf {\bibinfo {volume}
  {109}},\ \bibinfo {pages} {113105} (\bibinfo {year} {2016})}\BibitemShut
  {NoStop}%
\bibitem [{\citenamefont {Herner}\ \emph {et~al.}(2017)\citenamefont {Herner},
  \citenamefont {Weerakkody}, \citenamefont {Belkadi},\ and\ \citenamefont
  {Moddel}}]{Herner2017}%
  \BibitemOpen
  \bibfield  {author} {\bibinfo {author} {\bibfnamefont {S.~B.}\ \bibnamefont
  {Herner}}, \bibinfo {author} {\bibfnamefont {A.~D.}\ \bibnamefont
  {Weerakkody}}, \bibinfo {author} {\bibfnamefont {A.}~\bibnamefont {Belkadi}},
  \ and\ \bibinfo {author} {\bibfnamefont {G.}~\bibnamefont {Moddel}},\ }\href
  {\doibase 10.1063/1.4984278} {\bibfield  {journal} {\bibinfo  {journal}
  {Applied Physics Letters}\ }\textbf {\bibinfo {volume} {110}},\ \bibinfo
  {pages} {223901} (\bibinfo {year} {2017})}\BibitemShut {NoStop}%
\bibitem [{\citenamefont {Anderson}\ and\ \citenamefont
  {Cola}(2019)}]{Anderson2019}%
  \BibitemOpen
  \bibfield  {author} {\bibinfo {author} {\bibfnamefont {E.~C.}\ \bibnamefont
  {Anderson}}\ and\ \bibinfo {author} {\bibfnamefont {B.~A.}\ \bibnamefont
  {Cola}},\ }\href {\doibase 10.1021/acsaelm.9b00058} {\bibfield  {journal}
  {\bibinfo  {journal} {ACS Applied Electronic Materials}\ }\textbf {\bibinfo
  {volume} {1}},\ \bibinfo {pages} {692} (\bibinfo {year} {2019})}\BibitemShut
  {NoStop}%
\bibitem [{\citenamefont {Davids}\ \emph {et~al.}(2020)\citenamefont {Davids},
  \citenamefont {Kirsch}, \citenamefont {Starbuck}, \citenamefont {Jarecki},
  \citenamefont {Shank},\ and\ \citenamefont
  {Peters}}]{doi:10.1126/science.aba2089}%
  \BibitemOpen
  \bibfield  {author} {\bibinfo {author} {\bibfnamefont {P.~S.}\ \bibnamefont
  {Davids}}, \bibinfo {author} {\bibfnamefont {J.}~\bibnamefont {Kirsch}},
  \bibinfo {author} {\bibfnamefont {A.}~\bibnamefont {Starbuck}}, \bibinfo
  {author} {\bibfnamefont {R.}~\bibnamefont {Jarecki}}, \bibinfo {author}
  {\bibfnamefont {J.}~\bibnamefont {Shank}}, \ and\ \bibinfo {author}
  {\bibfnamefont {D.}~\bibnamefont {Peters}},\ }\href {\doibase
  10.1126/science.aba2089} {\bibfield  {journal} {\bibinfo  {journal}
  {Science}\ }\textbf {\bibinfo {volume} {367}},\ \bibinfo {pages} {1341}
  (\bibinfo {year} {2020})}\BibitemShut {NoStop}%
\bibitem [{\citenamefont {Elsharabasy}\ \emph {et~al.}(2020)\citenamefont
  {Elsharabasy}, \citenamefont {Bakr},\ and\ \citenamefont
  {Deen}}]{Elsharabasy2020}%
  \BibitemOpen
  \bibfield  {author} {\bibinfo {author} {\bibfnamefont {A.}~\bibnamefont
  {Elsharabasy}}, \bibinfo {author} {\bibfnamefont {M.}~\bibnamefont {Bakr}}, \
  and\ \bibinfo {author} {\bibfnamefont {M.~J.}\ \bibnamefont {Deen}},\ }\href
  {\doibase 10.1038/s41598-020-73368-7} {\bibfield  {journal} {\bibinfo
  {journal} {Scientific Reports}\ }\textbf {\bibinfo {volume} {10}},\ \bibinfo
  {pages} {16215} (\bibinfo {year} {2020})}\BibitemShut {NoStop}%
\bibitem [{\citenamefont {Periasamy}\ \emph {et~al.}(2013)\citenamefont
  {Periasamy}, \citenamefont {Guthrey}, \citenamefont {Abdulagatov},
  \citenamefont {Ndione}, \citenamefont {Berry}, \citenamefont {Ginley},
  \citenamefont {George}, \citenamefont {Parilla},\ and\ \citenamefont
  {O'Hayre}}]{Periasamy2013}%
  \BibitemOpen
  \bibfield  {author} {\bibinfo {author} {\bibfnamefont {P.}~\bibnamefont
  {Periasamy}}, \bibinfo {author} {\bibfnamefont {H.~L.}\ \bibnamefont
  {Guthrey}}, \bibinfo {author} {\bibfnamefont {A.~I.}\ \bibnamefont
  {Abdulagatov}}, \bibinfo {author} {\bibfnamefont {P.~F.}\ \bibnamefont
  {Ndione}}, \bibinfo {author} {\bibfnamefont {J.~J.}\ \bibnamefont {Berry}},
  \bibinfo {author} {\bibfnamefont {D.~S.}\ \bibnamefont {Ginley}}, \bibinfo
  {author} {\bibfnamefont {S.~M.}\ \bibnamefont {George}}, \bibinfo {author}
  {\bibfnamefont {P.~A.}\ \bibnamefont {Parilla}}, \ and\ \bibinfo {author}
  {\bibfnamefont {R.~P.}\ \bibnamefont {O'Hayre}},\ }\href {\doibase
  https://doi.org/10.1002/adma.201203075} {\bibfield  {journal} {\bibinfo
  {journal} {Advanced Materials}\ }\textbf {\bibinfo {volume} {25}},\ \bibinfo
  {pages} {1301} (\bibinfo {year} {2013})}\BibitemShut {NoStop}%
\bibitem [{\citenamefont {Herner}\ \emph {et~al.}(2018)\citenamefont {Herner},
  \citenamefont {Belkadi}, \citenamefont {Weerakkody}, \citenamefont {Pelz},\
  and\ \citenamefont {Moddel}}]{8272341}%
  \BibitemOpen
  \bibfield  {author} {\bibinfo {author} {\bibfnamefont {S.~B.}\ \bibnamefont
  {Herner}}, \bibinfo {author} {\bibfnamefont {A.}~\bibnamefont {Belkadi}},
  \bibinfo {author} {\bibfnamefont {A.}~\bibnamefont {Weerakkody}}, \bibinfo
  {author} {\bibfnamefont {B.}~\bibnamefont {Pelz}}, \ and\ \bibinfo {author}
  {\bibfnamefont {G.}~\bibnamefont {Moddel}},\ }\href {\doibase
  10.1109/JPHOTOV.2018.2791421} {\bibfield  {journal} {\bibinfo  {journal}
  {IEEE Journal of Photovoltaics}\ }\textbf {\bibinfo {volume} {8}},\ \bibinfo
  {pages} {499} (\bibinfo {year} {2018})}\BibitemShut {NoStop}%
\bibitem [{\citenamefont {Tekin}\ \emph {et~al.}(2021)\citenamefont {Tekin},
  \citenamefont {Weerakkody}, \citenamefont {Sedghi}, \citenamefont {Hall},
  \citenamefont {Werner}, \citenamefont {Wrench}, \citenamefont {Chalker},\
  and\ \citenamefont {Mitrovic}}]{Tekin2021}%
  \BibitemOpen
  \bibfield  {author} {\bibinfo {author} {\bibfnamefont {S.~B.}\ \bibnamefont
  {Tekin}}, \bibinfo {author} {\bibfnamefont {A.~D.}\ \bibnamefont
  {Weerakkody}}, \bibinfo {author} {\bibfnamefont {N.}~\bibnamefont {Sedghi}},
  \bibinfo {author} {\bibfnamefont {S.}~\bibnamefont {Hall}}, \bibinfo {author}
  {\bibfnamefont {M.}~\bibnamefont {Werner}}, \bibinfo {author} {\bibfnamefont
  {J.~S.}\ \bibnamefont {Wrench}}, \bibinfo {author} {\bibfnamefont {P.~R.}\
  \bibnamefont {Chalker}}, \ and\ \bibinfo {author} {\bibfnamefont {I.~Z.}\
  \bibnamefont {Mitrovic}},\ }\href {\doibase
  https://doi.org/10.1016/j.sse.2021.108096} {\bibfield  {journal} {\bibinfo
  {journal} {Solid-State Electronics}\ }\textbf {\bibinfo {volume} {185}},\
  \bibinfo {pages} {108096} (\bibinfo {year} {2021})}\BibitemShut {NoStop}%
\bibitem [{\citenamefont {Maraghechi}\ \emph {et~al.}(2012)\citenamefont
  {Maraghechi}, \citenamefont {Foroughi-Abari}, \citenamefont {Cadien},\ and\
  \citenamefont {Elezzabi}}]{Maraghechi2012}%
  \BibitemOpen
  \bibfield  {author} {\bibinfo {author} {\bibfnamefont {P.}~\bibnamefont
  {Maraghechi}}, \bibinfo {author} {\bibfnamefont {A.}~\bibnamefont
  {Foroughi-Abari}}, \bibinfo {author} {\bibfnamefont {K.}~\bibnamefont
  {Cadien}}, \ and\ \bibinfo {author} {\bibfnamefont {A.~Y.}\ \bibnamefont
  {Elezzabi}},\ }\href {\doibase 10.1063/1.3694024} {\bibfield  {journal}
  {\bibinfo  {journal} {Applied Physics Letters}\ }\textbf {\bibinfo {volume}
  {100}},\ \bibinfo {pages} {113503} (\bibinfo {year} {2012})}\BibitemShut
  {NoStop}%
\bibitem [{\citenamefont {Alimardani}\ and\ \citenamefont
  {Conley}(2013)}]{Alimardani2013}%
  \BibitemOpen
  \bibfield  {author} {\bibinfo {author} {\bibfnamefont {N.}~\bibnamefont
  {Alimardani}}\ and\ \bibinfo {author} {\bibfnamefont {J.~F.}\ \bibnamefont
  {Conley}},\ }\href {\doibase 10.1063/1.4799964} {\bibfield  {journal}
  {\bibinfo  {journal} {Applied Physics Letters}\ }\textbf {\bibinfo {volume}
  {102}},\ \bibinfo {pages} {1} (\bibinfo {year} {2013})}\BibitemShut {NoStop}%
\bibitem [{\citenamefont {Cui}\ \emph {et~al.}(2016)\citenamefont {Cui},
  \citenamefont {Sakhdari}, \citenamefont {Chamlagain}, \citenamefont {Chuang},
  \citenamefont {Liu}, \citenamefont {Cheng}, \citenamefont {Zhou},\ and\
  \citenamefont {Chen}}]{Cui2016}%
  \BibitemOpen
  \bibfield  {author} {\bibinfo {author} {\bibfnamefont {Q.}~\bibnamefont
  {Cui}}, \bibinfo {author} {\bibfnamefont {M.}~\bibnamefont {Sakhdari}},
  \bibinfo {author} {\bibfnamefont {B.}~\bibnamefont {Chamlagain}}, \bibinfo
  {author} {\bibfnamefont {H.-J.}\ \bibnamefont {Chuang}}, \bibinfo {author}
  {\bibfnamefont {Y.}~\bibnamefont {Liu}}, \bibinfo {author} {\bibfnamefont
  {M.~M.-C.}\ \bibnamefont {Cheng}}, \bibinfo {author} {\bibfnamefont
  {Z.}~\bibnamefont {Zhou}}, \ and\ \bibinfo {author} {\bibfnamefont {P.-Y.}\
  \bibnamefont {Chen}},\ }\href {\doibase 10.1021/acsami.6b11302} {\bibfield
  {journal} {\bibinfo  {journal} {ACS Applied Materials \& Interfaces}\
  }\textbf {\bibinfo {volume} {8}},\ \bibinfo {pages} {34552} (\bibinfo {year}
  {2016})}\BibitemShut {NoStop}%
\bibitem [{\citenamefont {Belkadi}\ \emph {et~al.}(2021)\citenamefont
  {Belkadi}, \citenamefont {Weerakkody},\ and\ \citenamefont
  {Moddel}}]{Belkadi2021}%
  \BibitemOpen
  \bibfield  {author} {\bibinfo {author} {\bibfnamefont {A.}~\bibnamefont
  {Belkadi}}, \bibinfo {author} {\bibfnamefont {A.}~\bibnamefont {Weerakkody}},
  \ and\ \bibinfo {author} {\bibfnamefont {G.}~\bibnamefont {Moddel}},\ }\href
  {\doibase 10.1038/s41467-021-23182-0} {\bibfield  {journal} {\bibinfo
  {journal} {Nature Communications}\ }\textbf {\bibinfo {volume} {12}},\
  \bibinfo {pages} {2925} (\bibinfo {year} {2021})}\BibitemShut {NoStop}%
\bibitem [{\citenamefont {Matsuura}\ \emph {et~al.}(2019)\citenamefont
  {Matsuura}, \citenamefont {Shimizu},\ and\ \citenamefont
  {Yugami}}]{Matsuura2019}%
  \BibitemOpen
  \bibfield  {author} {\bibinfo {author} {\bibfnamefont {D.}~\bibnamefont
  {Matsuura}}, \bibinfo {author} {\bibfnamefont {M.}~\bibnamefont {Shimizu}}, \
  and\ \bibinfo {author} {\bibfnamefont {H.}~\bibnamefont {Yugami}},\ }\href
  {\doibase 10.1038/s41598-019-55898-x} {\bibfield  {journal} {\bibinfo
  {journal} {Scientific Reports}\ }\textbf {\bibinfo {volume} {9}},\ \bibinfo
  {pages} {19639} (\bibinfo {year} {2019})}\BibitemShut {NoStop}%
\bibitem [{\citenamefont {Ward}\ \emph {et~al.}(2010)\citenamefont {Ward},
  \citenamefont {H{\"{U}}ser}, \citenamefont {Pauly}, \citenamefont {Cuevas},\
  and\ \citenamefont {Natelson}}]{Ward2010}%
  \BibitemOpen
  \bibfield  {author} {\bibinfo {author} {\bibfnamefont {D.~R.}\ \bibnamefont
  {Ward}}, \bibinfo {author} {\bibfnamefont {F.}~\bibnamefont {H{\"{U}}ser}},
  \bibinfo {author} {\bibfnamefont {F.}~\bibnamefont {Pauly}}, \bibinfo
  {author} {\bibfnamefont {J.~C.}\ \bibnamefont {Cuevas}}, \ and\ \bibinfo
  {author} {\bibfnamefont {D.}~\bibnamefont {Natelson}},\ }\href {\doibase
  10.1038/nnano.2010.176} {\bibfield  {journal} {\bibinfo  {journal} {Nature
  Nanotechnology}\ }\textbf {\bibinfo {volume} {5}},\ \bibinfo {pages} {732}
  (\bibinfo {year} {2010})}\BibitemShut {NoStop}%
\bibitem [{\citenamefont {Choi}\ \emph {et~al.}(2011)\citenamefont {Choi},
  \citenamefont {Yesilkoy}, \citenamefont {Ryu}, \citenamefont {Cho},
  \citenamefont {Goldsman}, \citenamefont {Dagenais},\ and\ \citenamefont
  {Peckerar}}]{Choi2011}%
  \BibitemOpen
  \bibfield  {author} {\bibinfo {author} {\bibfnamefont {K.}~\bibnamefont
  {Choi}}, \bibinfo {author} {\bibfnamefont {F.}~\bibnamefont {Yesilkoy}},
  \bibinfo {author} {\bibfnamefont {G.}~\bibnamefont {Ryu}}, \bibinfo {author}
  {\bibfnamefont {S.~H.}\ \bibnamefont {Cho}}, \bibinfo {author} {\bibfnamefont
  {N.}~\bibnamefont {Goldsman}}, \bibinfo {author} {\bibfnamefont
  {M.}~\bibnamefont {Dagenais}}, \ and\ \bibinfo {author} {\bibfnamefont
  {M.}~\bibnamefont {Peckerar}},\ }\href {\doibase 10.1109/TED.2011.2162414}
  {\bibfield  {journal} {\bibinfo  {journal} {IEEE Transactions on Electron
  Devices}\ }\textbf {\bibinfo {volume} {58}},\ \bibinfo {pages} {3519}
  (\bibinfo {year} {2011})}\BibitemShut {NoStop}%
\bibitem [{\citenamefont {Shin}\ \emph {et~al.}(2016)\citenamefont {Shin},
  \citenamefont {Im}, \citenamefont {Choi}, \citenamefont {Kim}, \citenamefont
  {Sohn}, \citenamefont {Cha},\ and\ \citenamefont {Jang}}]{Shin2016}%
  \BibitemOpen
  \bibfield  {author} {\bibinfo {author} {\bibfnamefont {J.~H.}\ \bibnamefont
  {Shin}}, \bibinfo {author} {\bibfnamefont {J.}~\bibnamefont {Im}}, \bibinfo
  {author} {\bibfnamefont {J.-W.}\ \bibnamefont {Choi}}, \bibinfo {author}
  {\bibfnamefont {H.~S.}\ \bibnamefont {Kim}}, \bibinfo {author} {\bibfnamefont
  {J.~I.}\ \bibnamefont {Sohn}}, \bibinfo {author} {\bibfnamefont {S.~N.}\
  \bibnamefont {Cha}}, \ and\ \bibinfo {author} {\bibfnamefont {J.~E.}\
  \bibnamefont {Jang}},\ }\href {\doibase
  https://doi.org/10.1016/j.carbon.2016.02.035} {\bibfield  {journal} {\bibinfo
   {journal} {Carbon}\ }\textbf {\bibinfo {volume} {102}},\ \bibinfo {pages}
  {172} (\bibinfo {year} {2016})}\BibitemShut {NoStop}%
\bibitem [{\citenamefont {Piltan}\ and\ \citenamefont
  {Sievenpiper}(2017)}]{Piltan2017}%
  \BibitemOpen
  \bibfield  {author} {\bibinfo {author} {\bibfnamefont {S.}~\bibnamefont
  {Piltan}}\ and\ \bibinfo {author} {\bibfnamefont {D.}~\bibnamefont
  {Sievenpiper}},\ }\href {\doibase 10.1063/1.4995995} {\bibfield  {journal}
  {\bibinfo  {journal} {Journal of Applied Physics}\ }\textbf {\bibinfo
  {volume} {122}},\ \bibinfo {pages} {183101} (\bibinfo {year}
  {2017})}\BibitemShut {NoStop}%
\bibitem [{\citenamefont {Fry-Bouriaux}\ \emph {et~al.}(2017)\citenamefont
  {Fry-Bouriaux}, \citenamefont {Rosamond}, \citenamefont {Williams},
  \citenamefont {Davies},\ and\ \citenamefont
  {W{\"{a}}lti}}]{PhysRevB.96.115435}%
  \BibitemOpen
  \bibfield  {author} {\bibinfo {author} {\bibfnamefont {L.}~\bibnamefont
  {Fry-Bouriaux}}, \bibinfo {author} {\bibfnamefont {M.~C.}\ \bibnamefont
  {Rosamond}}, \bibinfo {author} {\bibfnamefont {D.~A.}\ \bibnamefont
  {Williams}}, \bibinfo {author} {\bibfnamefont {A.~G.}\ \bibnamefont
  {Davies}}, \ and\ \bibinfo {author} {\bibfnamefont {C.}~\bibnamefont
  {W{\"{a}}lti}},\ }\href {\doibase 10.1103/PhysRevB.96.115435} {\bibfield
  {journal} {\bibinfo  {journal} {Phys. Rev. B}\ }\textbf {\bibinfo {volume}
  {96}},\ \bibinfo {pages} {115435} (\bibinfo {year} {2017})}\BibitemShut
  {NoStop}%
\bibitem [{\citenamefont {Shin}\ \emph {et~al.}(2017)\citenamefont {Shin},
  \citenamefont {Yang}, \citenamefont {Heo},\ and\ \citenamefont
  {Jang}}]{Shin2017}%
  \BibitemOpen
  \bibfield  {author} {\bibinfo {author} {\bibfnamefont {J.~H.}\ \bibnamefont
  {Shin}}, \bibinfo {author} {\bibfnamefont {J.~H.}\ \bibnamefont {Yang}},
  \bibinfo {author} {\bibfnamefont {S.~J.}\ \bibnamefont {Heo}}, \ and\
  \bibinfo {author} {\bibfnamefont {J.~E.}\ \bibnamefont {Jang}},\ }\href
  {\doibase 10.1063/1.5001149} {\bibfield  {journal} {\bibinfo  {journal} {AIP
  Advances}\ }\textbf {\bibinfo {volume} {7}},\ \bibinfo {pages} {105307}
  (\bibinfo {year} {2017})}\BibitemShut {NoStop}%
\bibitem [{\citenamefont {Gu}\ \emph {et~al.}(2006)\citenamefont {Gu},
  \citenamefont {Dey},\ and\ \citenamefont {Majhi}}]{Gu2006}%
  \BibitemOpen
  \bibfield  {author} {\bibinfo {author} {\bibfnamefont {D.}~\bibnamefont
  {Gu}}, \bibinfo {author} {\bibfnamefont {S.~K.}\ \bibnamefont {Dey}}, \ and\
  \bibinfo {author} {\bibfnamefont {P.}~\bibnamefont {Majhi}},\ }\href
  {\doibase 10.1063/1.2336718} {\bibfield  {journal} {\bibinfo  {journal}
  {Applied Physics Letters}\ }\textbf {\bibinfo {volume} {89}},\ \bibinfo
  {pages} {82907} (\bibinfo {year} {2006})}\BibitemShut {NoStop}%
\bibitem [{\citenamefont {Grover}\ and\ \citenamefont
  {Moddel}(2012)}]{GROVER201294}%
  \BibitemOpen
  \bibfield  {author} {\bibinfo {author} {\bibfnamefont {S.}~\bibnamefont
  {Grover}}\ and\ \bibinfo {author} {\bibfnamefont {G.}~\bibnamefont
  {Moddel}},\ }\href {\doibase https://doi.org/10.1016/j.sse.2011.09.004}
  {\bibfield  {journal} {\bibinfo  {journal} {Solid-State Electronics}\
  }\textbf {\bibinfo {volume} {67}},\ \bibinfo {pages} {94} (\bibinfo {year}
  {2012})}\BibitemShut {NoStop}%
\bibitem [{\citenamefont {Juppo}\ \emph {et~al.}(2000)\citenamefont {Juppo},
  \citenamefont {Rahtu}, \citenamefont {Ritala},\ and\ \citenamefont
  {Leskel{\"{a}}}}]{Juppo2000}%
  \BibitemOpen
  \bibfield  {author} {\bibinfo {author} {\bibfnamefont {M.}~\bibnamefont
  {Juppo}}, \bibinfo {author} {\bibfnamefont {A.}~\bibnamefont {Rahtu}},
  \bibinfo {author} {\bibfnamefont {M.}~\bibnamefont {Ritala}}, \ and\ \bibinfo
  {author} {\bibfnamefont {M.}~\bibnamefont {Leskel{\"{a}}}},\ }\href {\doibase
  10.1021/la991183+} {\bibfield  {journal} {\bibinfo  {journal} {Langmuir}\
  }\textbf {\bibinfo {volume} {16}},\ \bibinfo {pages} {4034} (\bibinfo {year}
  {2000})}\BibitemShut {NoStop}%
\bibitem [{\citenamefont {Elam}\ \emph {et~al.}(2003)\citenamefont {Elam},
  \citenamefont {Routkevitch}, \citenamefont {Mardilovich},\ and\ \citenamefont
  {George}}]{Elam2003}%
  \BibitemOpen
  \bibfield  {author} {\bibinfo {author} {\bibfnamefont {J.~W.}\ \bibnamefont
  {Elam}}, \bibinfo {author} {\bibfnamefont {D.}~\bibnamefont {Routkevitch}},
  \bibinfo {author} {\bibfnamefont {P.~P.}\ \bibnamefont {Mardilovich}}, \ and\
  \bibinfo {author} {\bibfnamefont {S.~M.}\ \bibnamefont {George}},\ }\href
  {\doibase 10.1021/cm0303080} {\bibfield  {journal} {\bibinfo  {journal}
  {Chemistry of Materials}\ }\textbf {\bibinfo {volume} {15}},\ \bibinfo
  {pages} {3507} (\bibinfo {year} {2003})}\BibitemShut {NoStop}%
\bibitem [{\citenamefont {Shrestha}\ \emph {et~al.}(2010)\citenamefont
  {Shrestha}, \citenamefont {Gu}, \citenamefont {Tran}, \citenamefont {Tapily},
  \citenamefont {Baumgart},\ and\ \citenamefont {Namkoong}}]{Shrestha_2010}%
  \BibitemOpen
  \bibfield  {author} {\bibinfo {author} {\bibfnamefont {P.}~\bibnamefont
  {Shrestha}}, \bibinfo {author} {\bibfnamefont {D.}~\bibnamefont {Gu}},
  \bibinfo {author} {\bibfnamefont {N.}~\bibnamefont {Tran}}, \bibinfo {author}
  {\bibfnamefont {K.}~\bibnamefont {Tapily}}, \bibinfo {author} {\bibfnamefont
  {H.}~\bibnamefont {Baumgart}}, \ and\ \bibinfo {author} {\bibfnamefont
  {G.}~\bibnamefont {Namkoong}},\ }\href {\doibase 10.1149/1.3485249}
  {\bibfield  {journal} {\bibinfo  {journal} {{ECS} Transactions}\ }\textbf
  {\bibinfo {volume} {33}},\ \bibinfo {pages} {127} (\bibinfo {year}
  {2010})}\BibitemShut {NoStop}%
\bibitem [{\citenamefont {Pyeon}\ \emph {et~al.}(2015)\citenamefont {Pyeon},
  \citenamefont {Cho}, \citenamefont {Baek}, \citenamefont {Kang},
  \citenamefont {Kim}, \citenamefont {Jeong},\ and\ \citenamefont
  {Kim}}]{Pyeon2015}%
  \BibitemOpen
  \bibfield  {author} {\bibinfo {author} {\bibfnamefont {J.~J.}\ \bibnamefont
  {Pyeon}}, \bibinfo {author} {\bibfnamefont {C.~J.}\ \bibnamefont {Cho}},
  \bibinfo {author} {\bibfnamefont {S.~H.}\ \bibnamefont {Baek}}, \bibinfo
  {author} {\bibfnamefont {C.~Y.}\ \bibnamefont {Kang}}, \bibinfo {author}
  {\bibfnamefont {J.~S.}\ \bibnamefont {Kim}}, \bibinfo {author} {\bibfnamefont
  {D.~S.}\ \bibnamefont {Jeong}}, \ and\ \bibinfo {author} {\bibfnamefont
  {S.~K.}\ \bibnamefont {Kim}},\ }\href@noop {} {\bibfield  {journal} {\bibinfo
   {journal} {Nanotechnology}\ }\textbf {\bibinfo {volume} {26}} (\bibinfo
  {year} {2015})}\BibitemShut {NoStop}%
\bibitem [{\citenamefont {Shimizu}\ \emph {et~al.}(2016)\citenamefont
  {Shimizu}, \citenamefont {Akutsu}, \citenamefont {Tsuda}, \citenamefont
  {Iguchi},\ and\ \citenamefont {Yugami}}]{Shimizu2016b}%
  \BibitemOpen
  \bibfield  {author} {\bibinfo {author} {\bibfnamefont {M.}~\bibnamefont
  {Shimizu}}, \bibinfo {author} {\bibfnamefont {H.}~\bibnamefont {Akutsu}},
  \bibinfo {author} {\bibfnamefont {S.}~\bibnamefont {Tsuda}}, \bibinfo
  {author} {\bibfnamefont {F.}~\bibnamefont {Iguchi}}, \ and\ \bibinfo {author}
  {\bibfnamefont {H.}~\bibnamefont {Yugami}},\ }\href
  {www.mdpi.com/journal/photonics} {\bibfield  {journal} {\bibinfo  {journal}
  {Photonics}\ }\textbf {\bibinfo {volume} {3}} (\bibinfo {year}
  {2016})}\BibitemShut {NoStop}%
\bibitem [{\citenamefont {Lee}\ \emph {et~al.}(2019)\citenamefont {Lee},
  \citenamefont {Bera}, \citenamefont {Shin}, \citenamefont {Hong},
  \citenamefont {Oh}, \citenamefont {Wan},\ and\ \citenamefont
  {Kwon}}]{https://doi.org/10.1002/admi.201901210}%
  \BibitemOpen
  \bibfield  {author} {\bibinfo {author} {\bibfnamefont {W.-J.}\ \bibnamefont
  {Lee}}, \bibinfo {author} {\bibfnamefont {S.}~\bibnamefont {Bera}}, \bibinfo
  {author} {\bibfnamefont {H.-C.}\ \bibnamefont {Shin}}, \bibinfo {author}
  {\bibfnamefont {W.-P.}\ \bibnamefont {Hong}}, \bibinfo {author}
  {\bibfnamefont {S.-J.}\ \bibnamefont {Oh}}, \bibinfo {author} {\bibfnamefont
  {Z.}~\bibnamefont {Wan}}, \ and\ \bibinfo {author} {\bibfnamefont {S.-H.}\
  \bibnamefont {Kwon}},\ }\href {\doibase
  https://doi.org/10.1002/admi.201901210} {\bibfield  {journal} {\bibinfo
  {journal} {Advanced Materials Interfaces}\ }\textbf {\bibinfo {volume} {6}},\
  \bibinfo {pages} {1901210} (\bibinfo {year} {2019})}\BibitemShut {NoStop}%
\bibitem [{\citenamefont {Gaillard}\ \emph {et~al.}(2006)\citenamefont
  {Gaillard}, \citenamefont {Pinzelli}, \citenamefont {Gros-Jean},\ and\
  \citenamefont {Bsiesy}}]{Gaillard2006}%
  \BibitemOpen
  \bibfield  {author} {\bibinfo {author} {\bibfnamefont {N.}~\bibnamefont
  {Gaillard}}, \bibinfo {author} {\bibfnamefont {L.}~\bibnamefont {Pinzelli}},
  \bibinfo {author} {\bibfnamefont {M.}~\bibnamefont {Gros-Jean}}, \ and\
  \bibinfo {author} {\bibfnamefont {A.}~\bibnamefont {Bsiesy}},\ }\href
  {\doibase 10.1063/1.2357891} {\bibfield  {journal} {\bibinfo  {journal}
  {Applied Physics Letters}\ }\textbf {\bibinfo {volume} {89}},\ \bibinfo
  {pages} {1} (\bibinfo {year} {2006})}\BibitemShut {NoStop}%
\end{thebibliography}%

\end{document}